\documentclass[twocolumn,showpacs,floatfix,amsmath,nofootinbib,amssymb,preprintnumbers,floatfix]{revtex4}
\usepackage{graphicx,color,dcolumn,booktabs,bm}
\usepackage{longtable,lscape}
\usepackage{txfonts}
\usepackage{overpic}
\usepackage{amssymb}
\usepackage{indentfirst}
\usepackage{feynmf}
\usepackage{slashed}
\usepackage{cases}
\usepackage{color}
\usepackage{multirow}
\usepackage{longtable}
\usepackage{graphicx,color,dcolumn,booktabs,bm}
\usepackage[colorlinks,
            citecolor=blue,
            anchorcolor=red,
            menucolor=red,
            linkcolor=red,
            filecolor=red,
            runcolor=red,
            urlcolor=blue,
            frenchlinks=red]{hyperref}

\begin{document}
\preprint{
\vbox{
\hbox{ADP-20-8/T1118}
}}

\title{Kaonic Hydrogen and Deuterium in Hamiltonian Effective Field Theory}

\author{Zhan-Wei Liu$^{1,5}$}\email{liuzhanwei@lzu.edu.cn}
\author{Jia-Jun Wu$^{4}$} 
\author{Derek B. Leinweber$^{2}$}
\author{Anthony W. Thomas$^{2,3}$}

\affiliation
{
$^1$School of Physical Science and Technology, Lanzhou University, Lanzhou 730000, China\\
$^2$Centre for the Subatomic Structure of Matter (CSSM), Department of Physics, University of Adelaide, Adelaide SA 5005, Australia\\
$^3$ARC Centre of Excellence for Particle Physics at the Terascale, Department of Physics, University of Adelaide, Adelaide SA 5005, Australia\\
$^4$School of Physical Sciences, University of Chinese Academy of Sciences (UCAS), Beijing
100049, China\\
$^5$Research Center for Hadron and CSR Physics, Lanzhou University and Institute of Modern Physics of CAS, Lanzhou 730000, China
}

\begin{abstract}
The anti-kaon nucleon scattering lengths resulting from a Hamiltonian effective field theory analysis of experimental data and lattice QCD studies are presented. The same Hamiltonian is then used to compute the scattering length for the $K^- d$ system, taking careful account of the effects of recoil on the energy at which the $\bar{K}N$ T-matrices are evaluated. These results are then used to estimate the shift and width of the $1S$ levels of anti-kaonic hydrogen and deuterium. The $K^- p$ result is in excellent agreement with the SIDDHARTA measurement. In the $K^- d$ case the imaginary part of the scattering length and consequently the width of the $1S$ state are considerably larger than found in earlier work. This is a consequence of the effect of recoil on the energy of the $\bar{K}N$ energy, which enhances the role of the $\Lambda(1405)$ resonance.

\end{abstract}

\maketitle

\section{Introduction}
\label{Introduction}
The kaon has played an important role in advancing particle physics since it was discovered in cosmic rays. The related $\theta$-$\tau$ puzzle resulted in the proposal of  parity violation. A bound state of the kaon and nucleon was predicted and the candidate $\Lambda(1405)$ was discovered in experiment, even before the birth of the quark model~\cite{Dalitz:1959dn,Alston:1961zzd}. The study of kaonic nuclei and atoms constitutes an important source of information as we seek to understand the nonperturbative properties of QCD and many experiments have been designed to investigate them (see, for example, Ref.~\cite{Gal:2016boi} for a recent review).  

In a  kaonic atom an electron is replaced by a $K^-$, which is still bound primarily by the attractive electromagnetic interaction. Since the kaon is three orders of magnitude heavier than the electron, the size of the kaonic orbit in a light atom is only $\sim$ 100 fm, while the binding energy is $\sim$ keV. If the kaon is captured by the nucleus, it can be annihilated in the formation of a hyperon. The behavior of these exotic atoms and nuclei allows us to explore new aspects of hadronic interactions. For example, the $\Lambda(1405)$ lies close below threshold in the $J^P=1/2^-$ $\bar KN$ system, while the lowest-lying $J^P=1/2^-$ strangeness-zero baryon, the $N(1535)$, is far from the $\pi N$ threshold. This difference has led to a great deal of speculation regarding the adequacy of the conventional quark model \cite{Veit:1984jr,Veit:1984an,Zhong:1988gn,Kaiser:1996js,Oset:1997it,Roca:2013cca,Ikeda:2011pi,Guo:2012vv,Liu:2016wxq,Xie:2013wfa,Mai:2014xna,Molina:2015uqp}.

Since the atomic size is beyond the range of the strong interaction, the kaonic atom can be studied pretty well within QED. However, for the $1S$ state, the wave function $\psi_{1S}^{QED}(\vec r=0)$ is non-zero  and the strong interaction plays a role. Therefore, the experimental energy, $E_{1S}$, is different from the QED prediction, $E_{1S}^{QED}$, because of the strong interaction, $\Delta E_{1S}\equiv E_{1S}-E_{1S}^{QED}$. Moreover, the $\bar K N$ system can decay to $\pi \Sigma$ and $\pi \Lambda$ and thus the $1S$ level has a width $\Gamma_{1S}$, in addition to the energy shift $\epsilon_{1S}$, $\Delta E_{1S}\equiv E_{1S}-E_{1S}^{QED}=\epsilon_{1S}-\frac{i}{2}\Gamma_{1S}$. The energy distributions of X-rays emitted from the excited kaonic atom may be measured to obtain $\Delta E_{1S}$ in experiment~\cite{Bazzi:2011zj,Bazzi:2012eq,Curceanu:2013bxa,Zmeskal:2019ksw,Zmeskal:2015efj,Curceanu:2019uph}. The difference between the measured and QED-predicted $nP-1S$ transition energy is essentially the same as $\Delta E_{1S}$, since the $nP$ level is scarcely affected by the strong interaction.
The SIDDHARTA experiment gave the energy shift and width for $1S$ kaonic hydrogen as 
\begin{eqnarray}
\epsilon_{1S}^p&=&283\pm 36({\rm stat})\pm 6({\rm sys})~{\rm eV},\nonumber\\
\Gamma_{1S}^p&=&541\pm 89({\rm stat})\pm 22({\rm sys})~{\rm eV} \, ,
\label{eq:KpExp}
\end{eqnarray}
which helps us to constrain the $K^- p$ interaction at low energy~\cite{Bazzi:2011zj,Bazzi:2012eq}. 
The reduced mass of the anti-kaon and the deuteron is a little larger than that of $\bar K N$ and thus kaonic deuterium would be easier to form under the pure electromagnetic interaction. There has been a proposal to search for kaonic deuterium in SIDDHARTA-2 and 
the J-PARC E57 experiment~\cite{Curceanu:2013bxa,Zmeskal:2019ksw,Zmeskal:2015efj,Curceanu:2019uph}. However, technically it is far more difficult to measure the kaonic deuteron than kaonic hydrogen. 

The study of kaonic deuterium may be expected to benefit from a comparison with earlier studies of pionic deuterium. The latter was proposed and observed in experiment very early -- see Ref.~\cite{Thomas:1979xu} for a review. The energy shifts and widths of the $1S$ states of pionic hydrogen and deuterium were measured with precision at the Paul Scherrer Institute (PSI) at Villigen~\cite{Hauser:1998yd}. 

Kaonic deuterium has already been investigated by many authors~\cite{Meissner:2006gx,Barrett:1999cw,Doleschall:2014wza,Ivanov:2004bv,Shevchenko:2014uva,Doring:2011xc,Mizutani:2012gy,Kamalov:2000iy,Hoshino:2017mty,Revai:2016muw,Sibirtsev:2006yw}. In recent studies, after the resummation of contributions from kaon multiple scatterings between the two nucleons in the deuteron, solving the full three-body Hamiltonian or Faddeev equations, the energy shift for kaonic deuterium has been found to be of order 700$\sim$900 eV, while  the width has been reported to be in the range 800$\sim$1200 eV~\cite{Doring:2011xc,Mizutani:2012gy,Kamalov:2000iy,Hoshino:2017mty,Revai:2016muw,Shevchenko:2014uva}.  

The nucleons in the deuteron have been assumed to be static in most studies of kaonic deuterium, perhaps because this approximation has been shown to be quite accurate for pionic deuterium. However, the situation merits further consideration for the following reason. In the pionic case the closest resonance, the $\Delta(1232)$, has angular momentum one and is still 150 MeV away from the $\pi N$ threshold. Even allowing for its width, it is far away. On the other hand, the difference between the $\Lambda(1405)$ resonance and the $\bar KN$ threshold is less than 30 MeV and, moreover, its 50 MeV width means they overlap to some extent. In such a case, even if the recoiling nucleon shifts the center-of-mass energy of $\bar KN$ a relatively small distance from the threshold, the scattering amplitude may be very different from that at threshold, which the static approximation uses. These considerations led us to examine the effect of recoil carefully in the $\bar K d$ system.

Some earlier studies of the effect of nucleon recoil for $\bar K d$ scattering was undertaken within effective field theory in Refs.~\cite{Baru:2009tx,Mai:2014uma}. There non-local effects were taken into account perturbatively by using the effective-range expansion of the $\bar KN$ amplitude.  The recoil effect for the double scattering contribution was found to be the order of 10-15\%, as compared to the static term \cite{Baru:2009tx}. The effect on the single scattering process was not discussed there, while it is a key issue in this study. 

In this work, we focus particularly on the recoil effects for the $1S$ level shifts of anti-kaonic deuterium with Hamiltonian Effective Field Theory (HEFT). HEFT was developed to study the low-lying resonances based on a combined analysis of  both lattice QCD data and experimental scattering data, while preserving the constraints of chiral perturbation theory where appropriate. It has led to important insights into the properties of the $N(1535)$, $N(1440)$, $\Lambda(1405)$, and so on~\cite{Hall:2013qba,Hall:2014uca,Liu:2016wxq,Liu:2016uzk,Liu:2015ktc,Wu:2017qve,Wu:2014vma,Leinweber:2015kyz,Liu:2017wsg}. In Ref.~\cite{Liu:2016wxq}, we studied both the cross sections for $K^- p$ scattering and the relevant lattice QCD data and reproduced the two-pole structure of the $\Lambda(1405)$. If not explicitly pointed out, references of the $\Lambda(1405)$ refer to the pole close to the experimental mass.

We first briefly review the HEFT for $\bar K N$ in Sec. \ref{sec:KN}. Then we use it to compute the $K^-p$ scattering lengths and the corresponding energy shift $\Delta E_{1S}^p$, with no further adjustment of the parameters determined in that previous work~\cite{Liu:2016wxq}. The scattering length of the $K^-d$ system and the energy shift $\Delta E_{1S}^d$ for kaonic deuterium, with and without the recoil correction, are presented in Secs. \ref{sec:SNA} and \ref{sec:REKD}.  We conclude with a discussion and summary in Sec. \ref{sec:NRD}. 
\begin{figure*}[tb]
\includegraphics[scale=0.9]{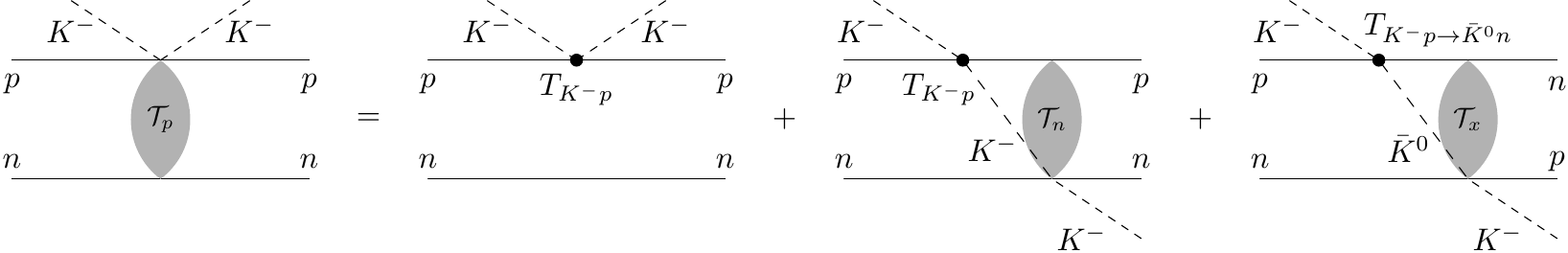}
\caption{Illustration of the Faddeev equation for $\mathcal T_p$ in kaon-deuteron scattering. The $T_{\bar KN}$ can be obtained from the study of kaon nucleon scattering, which contains the coupled-channel effect from $\bar KN$, $\pi\Sigma$, $\pi\Lambda$, and $\eta\Lambda$. There are two other similar diagrams for equations on $\mathcal T_n$ and $\mathcal T_x$ which can be found in Ref. \cite{Kamalov:2000iy}.}\label{Fig:Faddeev}
\end{figure*}
%
\section{Kaonic hydrogen and HEFT}\label{sec:KN}
The $T$ matrix for the two-body scatting process $\alpha\to\beta$ is related to the corresponding $S$ matrix and cross section by
\begin{eqnarray}
S_{\beta\alpha}&=&\delta_{\alpha\beta}-2i\sqrt{\rho_\alpha\rho_\beta}T_{\beta\alpha},\\
\sigma_{\beta\alpha}&=&\frac{4\pi^3 k^{\rm cm}_\beta\omega^{\rm cm}_{\alpha_M}\omega^{\rm cm}_{\alpha_B}\omega^{\rm cm}_{\beta_M}\omega^{\rm cm}_{\beta_B}}{k^{\rm cm}_\alpha(E_{\rm cm})^2}T_{\beta\alpha}T_{\beta\alpha}^*\label{eq_sigma} \, ,
\end{eqnarray}
where
\begin{equation}
\rho_\alpha=\pi \frac{\omega^{\rm cm}_{\alpha_M}\omega^{\rm cm}_{\alpha_B}}{E_{\rm cm}}k^{\rm cm}_\alpha \, .
\end{equation}
In this work, the channel label, $\alpha$ or $\beta$, refers to $\bar KN$, $\pi\Sigma$, $\pi\Lambda$, and $\eta\Lambda$ but, of course, it could be extended to more general cases. The labels $\alpha_M$ and $\alpha_B$ refer to the meson and baryon in channel $\alpha$, respectively. $\omega^{\rm cm}_{\alpha_X}$ and $k^{\rm cm}_\alpha$ denote the energy and spatial momentum of the particle in the center-of-mass frame, and $E_{\rm cm}$ is the total energy in the center-of-mass frame. 

For the process $\alpha\to \alpha$, the cross section $\sigma_{\alpha\alpha}^{\rm thr}$ at threshold can be related to the scattering length $a_\alpha$
\begin{equation}
\sigma_{\alpha\alpha}^{\rm thr}=4\pi |a_\alpha|^2,  \label{eq_sigmaA}
\end{equation}
where the scattering length is defined as 
\begin{equation}
a_\alpha\equiv-\lim_{k^{\rm cm}_\alpha\to 0} \pi \frac{\omega^{\rm cm}_{\alpha_M}\omega^{\rm cm}_{\alpha_B}}{E_{\rm cm}}T_{\alpha\alpha}
=- \pi\mu_\alpha T_{\alpha}^{\rm thr} \, 
\label{eq_slDef}
\end{equation}
and $\mu_\alpha$ is the reduced mass for the channel $\alpha$.

We used HEFT to study both the cross sections of $K^-p\to K^-p/\bar K^0 n/\pi^-\Sigma^+/\pi^0\Sigma^0/\pi^+\Sigma^-/\pi^0\Lambda$ and the corresponding energy levels in a finite volume (for comparison with lattice QCD simulations). The results were consistent with both the experimental scattering data and the spectra from lattice QCD simulations~\cite{Liu:2016wxq}. Here we want to further study the scattering length, $a_{K^-p}$, within this framework. Although there are two different scenarios which give different behavior at larger pion masses in Ref.~\cite{Liu:2016wxq,Menadue:2011pd}, we have checked that both yield the same threshold observables with negligible numerical differences. Thus we can choose the simpler scenario for this work. In our framework, the $T$ matrix can be obtained by solving the Bethe-Salpeter equation 
\begin{eqnarray}
&&T^I_{\alpha \beta}(k,k';E)=V^I_{\alpha \beta}(k,k')+\sum_\gamma \int q^2 \, dq\nonumber\\
&&\quad V^I_{\alpha \gamma}(k,q) \, \frac{1}{E-\omega_\gamma(q)+i \epsilon} \,  T^I_{\gamma \beta}(q,k';E),
\label{eq:BS}
\end{eqnarray}
with the Weinberg-Tomozawa potentials
\begin{equation}
V^I_{\alpha\beta}(k,k')=g_{\alpha,\beta}^I\frac{ \left [\, \omega_{\alpha_M}(k)+\omega_{\beta_M}(k')\,\right] \, u(k) \,u(k')}{8\pi^2 f^2 \,\sqrt{2\omega_{\alpha_M}(k)} \, \sqrt{2\omega_{\beta_M}(k')}} \, , \label{eq:vsp}
\end{equation}
where $f$ is the decay constant of the pion, the isospin $I$ can be 0 or 1. $k$ and $k'$ are the center-of-mass spatial momenta in the final and initial states, respectively. The form factor $u(k)=(1+k^2/\Lambda^2)^{-2}$ is used, with $\Lambda=1$ GeV \cite{Liu:2016wxq}.

To fix the nonzero couplings, $g_{\alpha,\beta}^I$, in Ref.~\cite{Liu:2016wxq} we fitted the cross sections for $K^-p\to K^-p/\bar K^0 n/\pi^-\Sigma^+/\pi^0\Sigma^0/\pi^+\Sigma^-/\pi^0\Lambda$ with $\chi_{d.o.f}=1.5$ for laboratory energies of the kaon up to 250 MeV. There are two poles for $\Lambda(1405)$ at $1428-23 \, i$ and $1338-89 \, i$ MeV with this set of parameters. We also compared our energy levels with lattice QCD results in Ref.~\cite{Liu:2016wxq}. 

We have not fine tuned the coupling constants, $g_{\alpha,\beta}^I$, in this work to exactly fit the scattering length. Instead, we use the values of $g_{\alpha,\beta}^I$ found in Ref.~\cite{Liu:2016wxq} to predict $a_{K^-p}$ and further check the reliability of HEFT.
The scattering lengths for $\bar K N$ in the isospin $I=0$ and $I=1$ channels in the HEFT, using Eq.~(\ref{eq_slDef}), are
\begin{equation}
a_{I=0}=-1.77+1.08\, i~~{\rm fm},\qquad
a_{I=1}=0.27+0.52\, i~~{\rm fm} \, . \label{eq:aI}
\end{equation}
With the isospin relation, one obtains
\begin{equation}
a_{K^-p}=-0.75+0.80 \, i~~{\rm fm},
\end{equation} 
and
\begin{eqnarray}
&&a_{K^- n}=0.27+0.52\, i~~{\rm fm},\qquad
a_{\bar K^0 n}=-0.75+0.80 \, i~~{\rm fm},\nonumber \\ 
&&a_{\bar K^0 n\to K^- p}=1.02-0.28 \, i~~{\rm fm} \, . \label{eq:aKNn}
\end{eqnarray} 

The relation between the energy shift $\epsilon_{1S}^p$ and width $\Gamma_{1S}^p$ of kaonic hydrogen and $a_{K^-p}$ can be given by the improved Deser formula~\cite{Meissner:2004jr}, but the logarithmic contribution at higher orders is important for kaonic deuterium~\cite{Hoshino:2017mty} and thus throughout this work we use the ``double-improved'' Deser formula in which the logarithmic terms are summed to all orders~\cite{Baru:2009tx}
\begin{equation}
\epsilon_{1S}^p-\frac{i}{2}\Gamma_{1S}^p=\frac{-2\alpha_e^3\,\mu_{K^-p}^2\, a_{K^-p}}{1+2\alpha_e\,\mu_{K^-p}\,(\ln \alpha_e-1)\,a_{K^-p}} \, , 
\label{eq:rel-rn}
\end{equation}
where $\alpha_e$ is the electromagnetic fine-structure constant.
With this resummed formula we obtain
\begin{equation}
\epsilon_{1S}^p=307~{\rm eV},\qquad
\Gamma_{1S}^p=533~{\rm eV} \, . 
\end{equation}
These results are  
consistent with the result of the SIDDHARTA experiment given  in Eq.~(\ref{eq:KpExp}). 

\section{results for kaonic deuterium with the static nucleon approximation}\label{sec:SNA}
The scattering amplitude $T_{\bar K d}$  can be divided into two parts
\begin{equation}
T_{\bar K d}=\int d^3 \vec q\,\psi_d(\vec q)\, \int d^3 \vec q'  \psi_d^*(\vec q')\,\,\,
\left(\mathcal T_p+\mathcal T_n\right),
\end{equation}
where $\psi_d(\vec q)$ is the wave function of the nucleon in the deuteron in momentum space  and $\vec q$ ($\vec q'$) is the relative momentum between the two nucleons in the initial (final) state. The $\bar K NN$ scattering amplitude $\mathcal T$ can be obtained by solving the Faddeev equations~\cite{Thomas:1977aab} illustrated in Fig.~\ref{Fig:Faddeev} 
\begin{eqnarray}
\mathcal T_p&\sim&T_{K^-p}\delta^3(\vec q-\vec q')+\int_{\delta}T_{K^-p}G_0\mathcal T_n+\int_{\delta}T_{K^-p\to \bar K^0 n} G_0\mathcal T_x,\nonumber\\
\mathcal T_n&\sim&T_{K^-n}\delta^3(\vec q-\vec q')+\int_{\delta}T_{K^-n}G_0\mathcal T_p,\nonumber\\
\mathcal T_x&\sim&T_{\bar K^0 n\to K^-p}\delta^3(\vec q-\vec q')+\int_{\delta}T_{\bar K^0 n}G_0\mathcal T_x+\int_{\delta}T_{\bar K^0 n\to K^-p} G_0\mathcal T_n \, .\nonumber\\
\label{eq:Faddeev}
\end{eqnarray}
It corresponds to the scattering length calculation, where the kaon momentum is zero and hence the momentum of the spectator nucleon (to the kaon-nucleon scattering) is its momentum in the deuteron center of mass. $G_0$ is the propagator 
\begin{equation}
G_0=(\epsilon_d-\frac{\vec p_K^2}{2m_K}-\frac{\vec p_{N_1}^2}{2m_N}-\frac{\vec p_{N_2}^2}{2m_N}+i0^+)^{-1}
,
\end{equation}
where $\vec p_K$, $\vec p_{N_1}$, and $\vec p_{N_2}$ are the momenta of the intermediate particles, and $\epsilon_d$ is the binding energy of deuteron.

The first term in Eq.~(\ref{eq:Faddeev}) corresponds to the single scattering (SS) of the kaon with only one nucleon, which contains the effects of the coupling of $\bar KN$ to the channels 
$\pi\Sigma$, $\pi\Lambda$, and $\eta\Lambda$. The remainder is generated by the multiple scattering (MS) of the kaon between the two nucleons in the deuteron.

If one uses the static nucleon approximation, usually on the basis that the momenta are relatively small, the momentum-dependent amplitude $T_{\bar K N}$ can be approximated by the value at threshold, $T_{\bar K N}^{\rm thr}$. With this approximation, the multiple scattering series integration can be solved analytically  and the scattering length, $a_{K^-d}$, can be simplified as~\cite{Kamalov:2000iy}
\begin{equation}
a_{K^-d}=\frac{m_d}{m_K+m_d}\int d^3 \vec r \,|\psi_d(\vec r)|^2\,\, \hat A_{K^-d}(r) \, ,
\label{eq_akd_static}
\end{equation}
where $\psi_d(\vec r)$ is the wave function for nucleons in the deuteron, and 
\begin{equation}
\hat A_{K^-d}(r)=\frac{\tilde a_{K^-p}+\tilde a_{K^-n}+(2\tilde a_{K^-p}\tilde a_{K^-n}-b_x^2)/r-2b_x^2\tilde a_{K^-n}/r^2}{1-\tilde a_{K^-p}\tilde a_{K^-n}/r^2+b_x^2\tilde a_{K^-n}/r^3} \, , \label{eq:aKd}
\end{equation} 
with $\tilde a_{\bar K N}=a_{\bar K N}(1+m_K/m_N)$, and\\ 
$b_x= \tilde a_{\bar K^0 n\to K^- p}/\sqrt{1+\tilde a_{\bar K^0 n}/r}$. This corresponds to the commonly used  fixed center approximation (e.g. Ref.~\cite{Roca:2010tf}) in which excited nucleon intermediate states are neglected.

Using the $\bar KN$ scattering lengths given in Eq.~(\ref{eq:aKNn}) and the deuteron wave function obtained from a one-boson-exchange model~\cite{Chen:2017jjn}, the numerical result with the static nucleon approximation is
\begin{equation}
a_{K^-d}|_{\rm StaticApprox}=-1.55+1.57\, i~~{\rm fm} \, .
\end{equation}
With the resummed formula (\ref{eq:rel-rn}), the energy shift and width of the $1S$ level of kaonic deuterium is
\begin{equation}
\epsilon_{1S}^d|_{\rm StaticApprox}=855~{\rm eV},\quad
\Gamma_{1S}^d|_{\rm StaticApprox}=1127~{\rm eV} \, . 
\label{eq:e_d2}
\end{equation}
%

\section{recoil effects for kaonic deuterium}\label{sec:REKD}
The root-mean-square radius of the deuteron is about 2 fm, which means the momentum $q$ of the nucleons in the deuteron should be of order 1/(2 fm) $\sim$ 100 MeV on average. The variation of momentum is not very large, but the resonance $\Lambda(1405)$ is around the $\bar KN$ threshold and thus the effect from the nucleon recoiling in the deuteron should be carefully examined. 

Firstly, we study the single scattering term of $K^- d$, which can be represented as the second diagram in 
Fig.~\ref{Fig:Faddeev}. Denoting the scattering amplitude of a static kaon and a nucleon with momentum $\vec q$ in the deuteron as $T_{\bar K N}(\vec q)$, the integrated amplitude is
\begin{equation}
\langle\,T_{\bar K N}^d\,\rangle\equiv \int d^3\vec q\, |\psi_d(\vec q)|^2\,\, T_{\bar K N}(\vec q). \label{eq:T_KNd}
\end{equation}  
Obviously, the amplitude of $K^-d$ from the single scattering contribution is exactly $T_{K^-d}^{SS}=\langle\,T_{K^- p}^d\,\rangle+\langle\, T_{K^- n}^d\,\rangle$. 

To calculate $T_{\bar K N}(\vec q)$, we need to recognise that if the spectator nucleon has momentum $\vec q$, the $\bar{K} N$ center of mass energy at which the two-body T-matrix must be evaluated is effectively $E_{\bar KN}=m_d + m_K - m_N-q^2/2\mu_3$ within Faddeev formalism where $\mu_3=m_N(m_N+m_K)/(2m_N+m_K)$. 
We use the following relation to obtain $T_{\bar K N}(\vec q)$
\begin{equation}
T_{\bar K N}(\vec q)=T_{\bar KN,\bar KN}(k,k;E_{\bar KN}) \, ,
\end{equation}
where $k=q m_K/(m_N+m_K)$.

For the multiple scattering terms, we use the approximation $T_{\bar K N}\to \langle\,T_{\bar K N}^d\,\rangle$ to simplify the calculation. 
Finally, we can evaluate $a_{K^-d}$. 
By replacing $a_{\bar K N}$ with $\langle\,a_{\bar K N}^d\,\rangle\equiv -\pi\mu_{\bar K N} \langle\,T_{\bar K N}^d\,\rangle$ in the formula given in Eq.~(\ref{eq_akd_static}), the $K^-d$ scattering length, including the recoil correction, may be obtained.

\section{numerical results and discussion}\label{sec:NRD}
More than 96\% of the contribution to $\langle\,T_{\bar K N}^d\,\rangle$ in Eq. (\ref{eq:T_KNd}) comes from $q < 300$ MeV, which corresponds to $k\lesssim 110$ MeV and $\Delta E_{\bar K N}\lesssim 80$ MeV, a region in which our fit with HEFT is very reliable~\cite{Liu:2016wxq}. The wave function $\psi_d(\vec q)$ is extracted from the Bonn potential \cite{ENegele:1989pd}. The wave function of the deuteron is well constrained both by experiments and theory, and thus we directly cite two different results from Refs. \cite{Chen:2017jjn} and \cite{ENegele:1989pd} for $\psi_d(\vec r)$ and $\psi_d(\vec q)$ in the coordinate and momentum spaces respectively. We do not transform between them and simply neglect the modest differences between the models.

\begin{figure}[tb]
\includegraphics[scale=0.5]{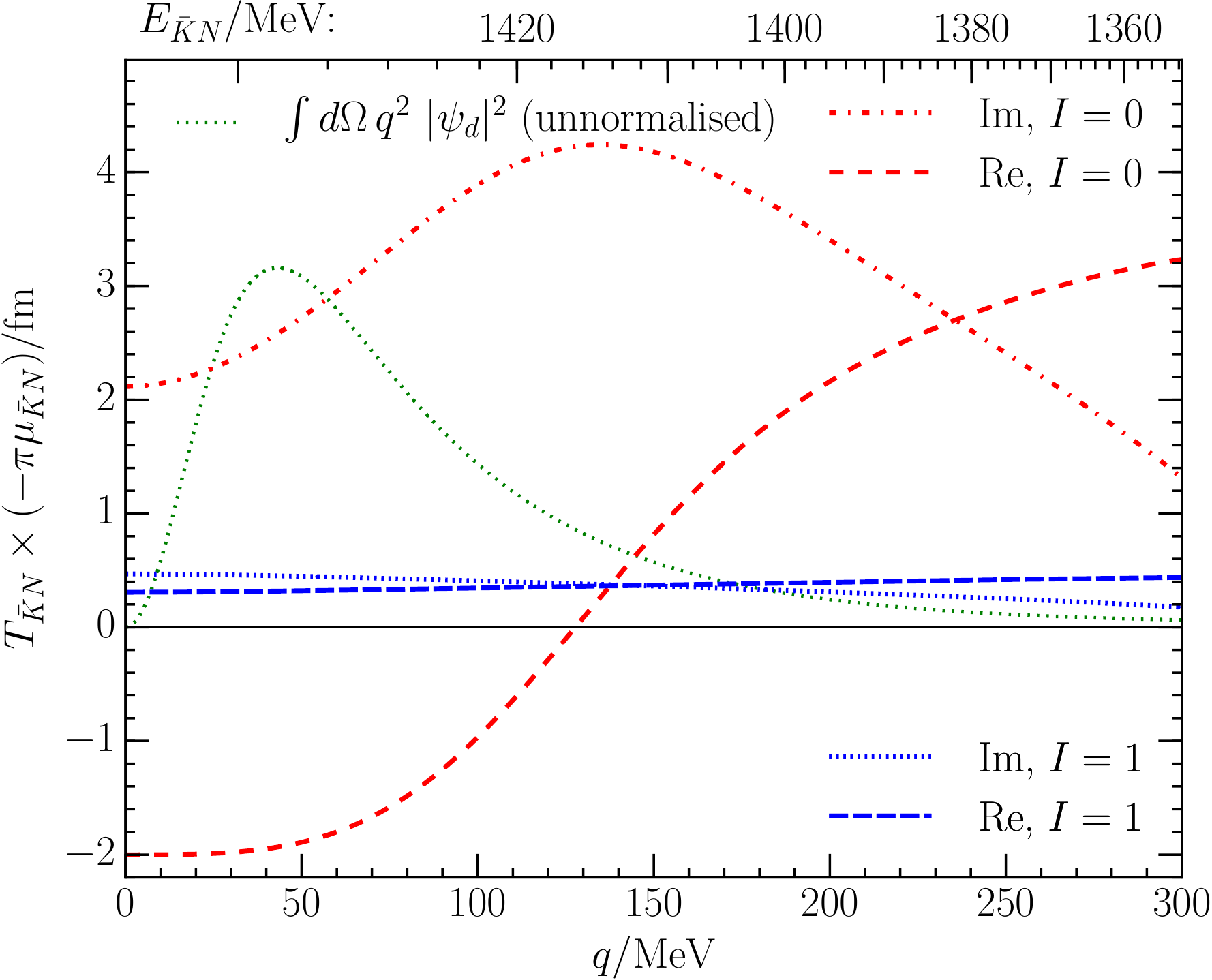}
\caption{The scattering amplitude $T_{\bar K N}(\vec q)$ for a static kaon and a nucleon of momentum $q$ in the deuteron. The lower x-axis is the spatial momentum $q$ of a nucleon, while the upper x-axis is the center-of-mass energy $E_{\bar K N}$ of $\bar K N$, and they are related by $E_{\bar K N}=(m_d/2+m_K)-q^2/2\mu_3$ with $\mu_3=m_N(m_N+m_K)/(2m_N+m_K)$. The deuteron wave function is also shown as the green dotted lines.}
\label{fig:TKNd}
\end{figure}

Before providing the average $\langle\,T_{\bar K N}^d\,\rangle$, we show the scattering amplitude $T_{\bar K N}(\vec q)$ in Fig.~\ref{fig:TKNd}. By comparing the two x-axes, one can see that even if the momentum $q$ increases from 0 to 200 MeV where the wave function $\psi_d$ is extremely close to 0, the center-of-mass energy $E_{\bar KN}$ of the $\bar K N$ system varies over a relatively small range, roughly 1430 to 1400 MeV. Usually such a small change of $E_{\bar KN}$ would not be expected to lead to large variations in related amplitudes. For example, the blue dashed lines for the amplitudes $T_{\bar K N}(I=1)$ are relatively flat for isospin 1. However, the issue is very different when there is a resonance or bound state in a channel below and close to the threshold.  
\begin{table*}[tb]
\caption{Comparison of scattering lengths $a_{K^-d}$ in units of fm. The first four columns are from Refs. \cite{Kamalov:2000iy,Doring:2011xc,Mizutani:2012gy,Hoshino:2017mty}. The last four  columns are the results in this work with (\checkmark) and without ($\times$) the multiple scattering contributions or the recoil effects, respectively.}
\label{tab:com}
\begin{tabular}{c|cccc|cccc}
$a_{K^-d}$ &Kamalov \cite{Kamalov:2000iy}&Doring \cite{Doring:2011xc}&Mizutani \cite{Mizutani:2012gy}&Hoshino \cite{Hoshino:2017mty}&(MS \checkmark, recoil $\times$)&(MS $\times$, recoil \checkmark)&(MS $\times$, recoil $\times$)&(MS \checkmark, recoil \checkmark)\\
\hline
Re&$-1.62$&$-1.46$&$-1.58$&$-1.42$&$-1.55$&-0.06&-0.58  &$-0.59$\\
Im&1.91&1.08&1.37&1.60& 1.57    &2.55&1.59 &2.70\\
\end{tabular}
\end{table*}

In our previous work \cite{Liu:2016wxq} we searched for two poles for the $\Lambda(1405)$ in the $I=0$ channel within HEFT, finding one close to the $\bar KN$ threshold with width about 50 MeV. As $E_{\bar KN}$ passes by the pole, the scattering amplitude $T_{\bar K N}(I=0)$ changes rapidly, as shown by the red dashed and dot-dashed lines in Fig.~\ref{fig:TKNd}. The phase shift usually crosses $90^{\tiny o}$ when a resonance appears, which is equivalently exhibited by the real part of the amplitude changing its sign for $E_{\bar K N}$ around 1410 MeV (red dashed). The imaginary part of $T_{\bar K N}(I=0)$ is related to the experimental $\pi\Sigma$ invariant mass distribution and thus the red dot-dashed line shows a bump \cite{Hemingway:1984pz}. 

From this analysis, we would expect that the average value of $\langle\,T_{\bar K N}^d\,\rangle$ may deviate significantly from the threshold value in the $I=0$ channel. Equivalently speaking, $\langle\,a_{I=0}^d\,\rangle$ is very different from the scattering length $a_{I=0}$, while $\langle\,a_{I=1}^d\,\rangle$ should be still close to $a_{I=1}$. We find 
\begin{equation}
\langle\,a_{I=0}^d\,\rangle=-1.07+3.01\, i~~{\rm fm},\qquad
\langle\,a_{I=1}^d\,\rangle=0.32+0.40\, i~~{\rm fm} \, ,
\end{equation}
which in comparison with the threshold scattering lengths in Eq.~(\ref{eq:aI}) shows that the average imaginary part in the channel with $I=0$ increases by more than 1 fm. This will, of course, increase the width of the $1S$ $K^- d$ energy level significantly.

We show the $K^-d$ scattering length after taking the recoiling effect into consideration
\begin{equation}
a_{K^-d}=-0.59+2.70\, i~~{\rm fm} \, .
\end{equation}
With the resummed formula of Eq. (\ref{eq:rel-rn}), the energy shift and width of $1S$ kaon deuterium are 
\begin{equation}
\epsilon_{1S}^d=803~{\rm eV},\qquad
\Gamma_{1S}^d=2280~{\rm eV}. \label{eq:e_d2full}
\end{equation}
By comparing these results with those given in Eq.~(\ref{eq:e_d2}) of Sec.~\ref{sec:SNA}, we can clearly see that the recoil effect plays a very important role for the $K^-d$ system.

\begin{figure}[t]
\includegraphics[scale=0.45]{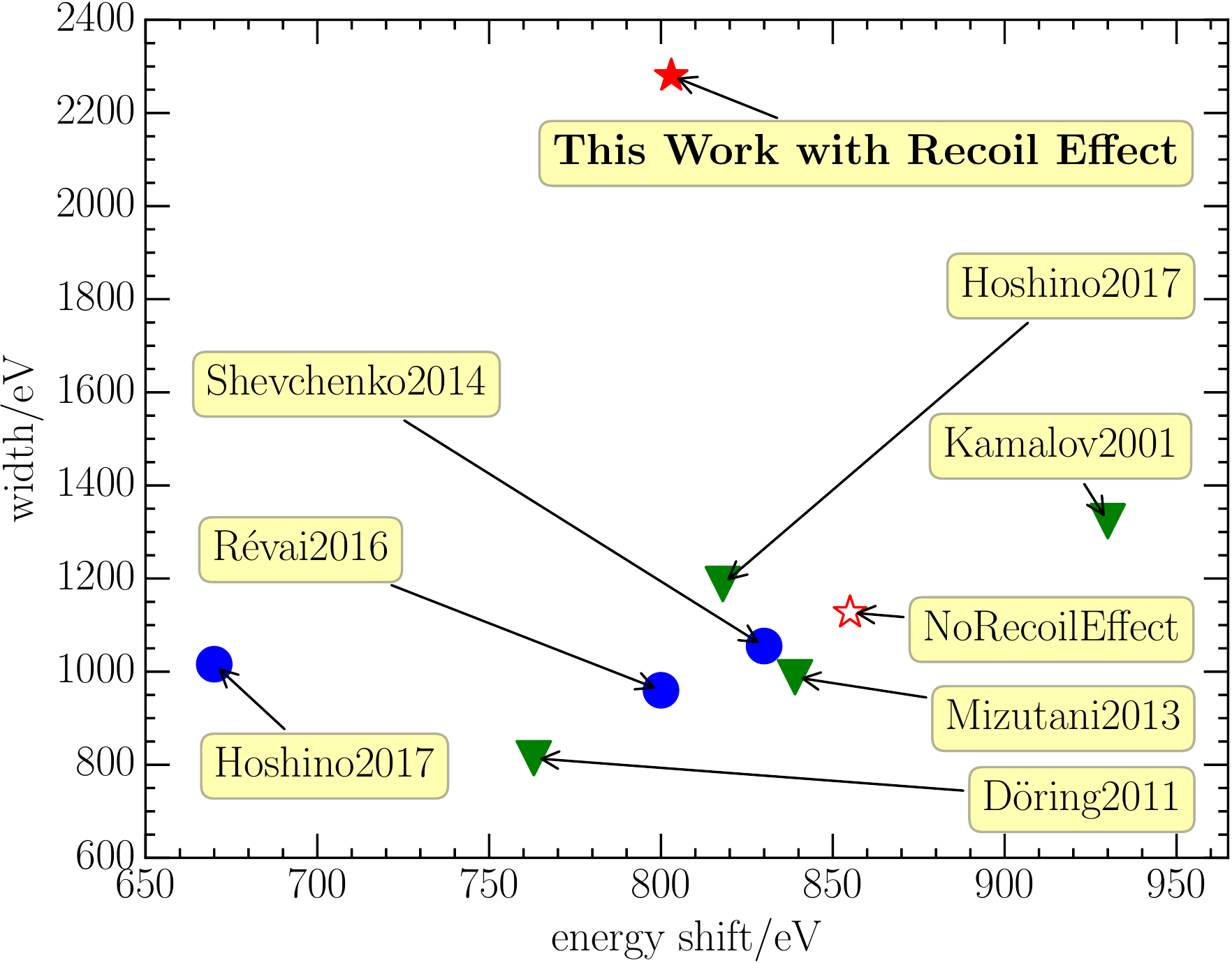}
\caption{Comparison of the energy shifts and widths for the $1S$ state of kaonic deuterium. The data with green triangles are from Refs. \cite{Doring:2011xc,Mizutani:2012gy,Kamalov:2000iy,Hoshino:2017mty} and $\Delta E_{1S}^d$ are extracted from the scattering lengths of $K^- d$ using the resummed formula (\ref{eq:rel-rn}). The data with blue circles are from Refs. \cite{Revai:2016muw,Shevchenko:2014uva,Hoshino:2017mty} where the dynamical equations directly provide the energy level of kaonic deuterium without the help of $K^-d$ scattering length. For example, there are two points labeled with ``Hoshino2017'' which are from the same paper \cite{Hoshino:2017mty}, but the left blue circle is directly obtained by solving the Schr\"odinger equation while the right green triangle is translated with the double-improved Deser formula (\ref{eq:rel-rn}) from the $K^-d$ scattering length. Our result (labeled with hollow star) without the recoiling effect is close to other works, but the result after considering the recoiling effect (labeled with a filled star) has a much larger width.}\label{fig:com}
\end{figure}
We compare the scattering lengths $a_{K^-d}$ in this work with some other studies in Table~\ref{tab:com} and 
Fig. \ref{fig:com}. Without including the effect of recoil  our results in the fifth column are close to those of earlier work. That is, our result would be close to others if the momentum of the nucleon in the deuteron is neglected. However, our corrected result has a much larger width.

The contribution to $a_{K^-d}$ from the single scattering term is exactly
\begin{equation}
a_{K^-d}^{SS}=\frac{\mu_{\bar K d}}{\mu_{\bar K N}}\left(\,\frac12 \langle\,a_{I=0}^d\,\rangle +\frac32\langle\,a_{I=1}^d\,\rangle\,\right).
\end{equation} 
\vspace{-0.5em}\\%
We divide $a_{K^-d}$ into $a_{K^-d}^{SS}$ and $a_{K^-d}^{MS}$, which are from the single and multiple scattering contributions, respectively
\begin{equation}
a_{K^-d}=a_{K^-d}^{SS}+a_{K^-d}^{MS}=(-0.06+2.55i)+(-0.53+0.15i)~{\rm fm} \, .
\end{equation}
By comparing this with  
\begin{equation}
a_{K^-d}|_{\rm StaticApprox}=(-0.58+1.59i)+(-0.97-0.02i)~{\rm fm} \, ,
\end{equation}
the main difference in the imaginary part clearly arises from the single scattering term. This gives us confidence in the conclusion that the ground state of kaonic deuterium is broad. Our final result for the $K^-d$ scattering length is
\begin{equation}
a_{K^-d}= \, -0.59 \, + \, 2.70 \, i~~{\rm fm}  \, .
\end{equation} 

In summary, we first established that our earlier Hamiltonian effective field theory study of the $\bar{K} p$ system did  indeed reproduce the empirical $K^-p$ scattering length, as well as producing an energy shift and width for the ground state of kaonic hydrogen which is consistent with the SIDDHARTA experiment. The calculation was then extended to the the $K^-d$ system, where it was found that the effect of recoil in the energy argument of the $\bar{K} p$ T-matrix within the Faddeev formalism is very significant, making the $1S$ level of kaonic deuterium considerably more broad and short lived.

\section*{Acknowledgement}
Z.W.L thanks Kan Chen for checking the program of kaon nucleon scattering length. This project is supported by the National Natural Science Foundation of China under Grants Nos. 11705072 and 11965016, CAS Interdisciplinary Innovation Team (Z.W.L), the Thousand Talents Plan for Young Professionals (J.J.W), and the Australian Research Council through ARC Discovery Project Grants Nos. DP150103101 and DP180100497 (A.W.T) and DP150103164 and DP190102215 (D.B.L).


\end{document}